# The effect of the dielectric end groups on the positive bias stress stability of N2200 organic field effect transistors

D. Simatos[1,b)], L. J. Spalek[1,a),b)], U. Kraft[1], M. Nikolka[1], X. Jiao[2], C. R. McNeill[2], D. Venkateshvaran[1,a)], H. Sirringhaus[1,a)]

**AFFILIATIONS**

[1]Cavendish Laboratory, University of Cambridge, J.J. Thomson Avenue, Cambridge CB3 0HE, UK

[2]Department of Materials Science and Engineering, Monash University, Wellington Road, Clayton, VIC 3800, Australia

a) Authors to whom correspondence should be addressed**:** ljs51@cam.ac.uk, dv246@cam.ac.uk, hs220@cam.ac.uk

b) D. Simatos and L. J. Spalek contributed equally to this work.

**ABSTRACT**

Bias stress degradation in conjugated polymer field-effect transistors is a fundamental problem in these disordered materials and can be traced back to interactions of the material with environmental species,[1,2,3] as well as fabrication-induced defects.[4,5] However, the effect of the end groups of the polymer gate dielectric and the associated dipole-induced disorder on bias stress stability has not been studied so far in high-performing n-type materials, such as N2200.[6,7] In this work, the performance metrics of N2200 transistors are examined with respect to dielectrics with different end groups (Cytop-M and Cytop-S[8]). We hypothesize that the polar end groups would lead to increased dipole-induced disorder, and worse performance.[1,9,10] The long-time annealing scheme at lower temperatures used in the paper is assumed to lead to better crystallization by allowing the crystalline domains to reorganize in the presence of the solvent.[11] It is hypothesized that the higher crystallinity could narrow down the range at which energy carriers are induced and thus decrease the gate dependence of the mobility. The results show that the dielectric end groups do not influence the bias stress stability of N2200 transistors. However, long annealing times result in a dramatic improvement in bias stress stability, with the most stable devices having a mobility that is only weakly dependent on or independent of gate voltage.

## I. INTRODUCTION

Electrical degradation during device operation (also known as bias stress) is a well-known phenomenon in organic semiconductors and other disordered materials. It manifests itself as an exponential decay of the output current under the continuous application of bias, a shift of the threshold voltage, a decrease of the charge carrier mobility, an increase of the sub-threshold slope, and the appearance of hysteresis.[1,2] To a large extent, the origin of the observed device deterioration stems from interactions between the organic semiconductor and light and atmospheric species such as oxygen or water. However, the overall stress behavior can also be impacted by other factors such as the choice of solvent used for processing,[11] or chemical impurities present in the materials or introduced during the fabrication and storage of the samples.[4] Many studies have traced the origins of bias stress instability to water adsorbed on the semiconductor/dielectric interface,[5] photo-induced degradation,[4] or the solvent-tuned crystallinity of the material.[11] Other studies have also examined the effects of the polarity of the dielectric, with the more polar dielectrics inducing dipole-induced energetic disorder.[1,9] The disorder may also originate by the dipole moments of the dielectric end groups.[10]

Despite extensive research on the origins and improvement of bias stress instability, to date there are very few studies that examine the effect of the dielectric end groups on bias stress stability. The end groups of the polymer chain can be used to tune the physical properties of a polymer dielectric, e.g., to induce cross-linking functionalities, but it is important to understand whether and how operational reliability might be affected by different end groups. Here we examine how the end groups of the commonly used fluoropolymer gate dielectric, Cytop, affect the bias stress stability of N2200, also known as P(NDI-2OD-T2),[6,7] which is one of the highest-performing, semi-crystalline n-type polymers. We focus on N2200 because environmental stability and bias stress stability are still open questions for n-type polymers.[7,12,13] To investigate the effect of end group induced dipolar disorder, we examine two variations of the dielectric: a silane-terminated (Cytop-M) and fluorine-terminated (Cytop-S) variant. There is also a carboxylic acid-terminated (Cytop-A) variant, which is not studied here, because we are examining whether the absence of any functional groups improves the device stability. Cytop-M's silane end groups are reactive with water, and the reaction byproducts can potentially compromise device performance.[14] The polar nature of the end groups is also expected to induce energetic disorder at the semiconductor-dielectric interface. In contrast, Cytop-S has a less reactive end group, and thus is expected to have no effect on device stability. Because we are interested in how the end groups affect the electronic structure and dipolar disorder at the interface we perform our bias stress investigations in an inert nitrogen atmosphere in order to eliminate the fast charge trapping that would occur in the presence of water/oxygen as much as possible.[13,15] Our results show that Cytop-S is a well-performing gate dielectric with comparable performance and operational stability to the widely used Cytop-M.

To investigate the role of end groups more systematically we explored a range of processing recipes that have been reported in the literature. We hypothesize that long-time annealing schemes at lower temperatures will allow the crystalline domains of the polymer to reorganize in the presence of the solvent, and lead to better crystallization. Improved crystallization has been linked to improved bias stress stability,[11] as greater crystallinity narrows down the range of energy levels at which carriers are induced. We found that long annealing times improve bias stress performance dramatically for both dielectrics, and that this improvement is not attributed to the crystallinity of the polymer. We observed an interesting correlation between the gate voltage dependence of the mobility when the devices were measured initially and the operational device stability during long-time bias stress testing. Devices with a strongly gate voltage dependent mobility were the most unstable, whereas those with a weaker gate voltage dependent or even gate voltage independent mobility were the most stable.

## II. DEVICE FABRICATION

Top gate bottom contact field effect transistors (FETs) with both grades of Cytop were fabricated as shown in **FIG. 1** and described in the Experimental Methods Section under different process conditions reported in the literature: After deposition by spin coating N2200 was annealed in two different ways: a short annealing time above the solvent's boiling point (200°C for 15 minutes), and a long annealing time below the boiling point (110°C for 14 hours). The annealing scheme with high temperature and short annealing time aims to remove the solvent entirely from the polymer film. Conversely, the long-time annealing scheme at lower temperatures aims to retain some solvent in the film, and allow the crystalline domains to reorganize, in a high boiling point solvent, to allow for better crystallization. The literature suggests that the long-term annealed film below the boiling point would have better crystallinity, and hence improved bias stress stability.[11]

For the dielectric, we started off by comparing two different annealing recipes found in the literature: one with a short annealing time (90°C for 20 minutes),[16] and one with a longer one (110°C for 5 hours),[17] both being below the Cytop solvent's boiling point (180°C). Additionally, to investigate the effect of the Cytop solvent on device performance, we also used pure Cytop.

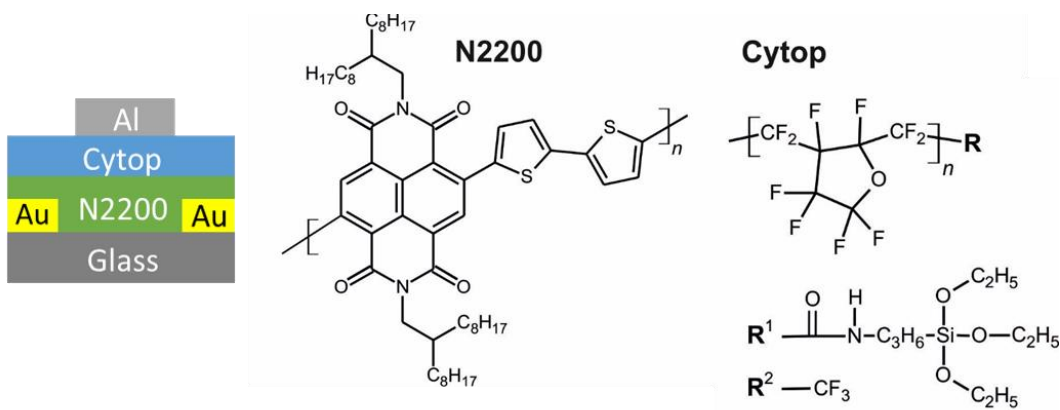

**FIG 1.** Device architecture. Top gate bottom contact (TGBC) device architectures were used, with N2200, also known as P(NDI-2OD-T2), as the organic semiconductor, and two different Cytop variants: Cytop-M, which is silane-terminated, and Cytop-S, which is fluorine-terminated.

To avoid metal penetration into the dielectric, we always evaporated the gate in a two-step recipe, with the first 10 nm evaporated with a small rate (0.2 A/sec). To investigate the effect of post-fabrication annealing, the devices were annealed after the gate was evaporated at 80°C for 14-15 hours.[18] The effect of this post-fabrication annealing step was investigated through the fabrication of a batch that was annealed at 80°C before the gate was evaporated. It is worth noting that in the case where N2200 and Cytop were annealed for a short time (recipe 1), the post-fabrication annealing step damaged the gates, which is assumed to be due to solvent leaving the Cytop film when the device was exposed to the vacuum of the evaporator (**FIG. S1**). **Table I** summarizes the fabrication details and different recipes that were compared for this study. **Table SI** compares the results of this study with those of the literature. A detailed discussion of the fabrication protocols is included in the Experimental Methods Section.

**Table I.** Fabrication details

| Recipe | OSC | Insulator | Gate |
|---|---|---|---|
| 1 | N2200 Pristine 7 g/L (100% DCB) annealed at 200°C for 15 minutes. | Cytop M-grade (3:1) annealed at 90°C for 20 minutes. Cytop S-grade (1.5:1) annealed at 90°C for 20 minutes. | Annealed at 80°C for 14 hours **after** gate evaporation. |
| 2 | N2200 Pristine 7 g/L (100% DCB) annealed at 200°C for 15 minutes. | Cytop M-grade (3:1) annealed at 90°C for 20 minutes. Cytop S-grade (1.5:1) annealed at 90°C for 20 minutes. | Annealed at 80°C for 14 hours **before** gate evaporation. |
| 3 | N2200 Pristine 7 g/L (100% DCB) annealed at 110°C overnight (14 hours). | Cytop M-grade 100% annealed at 110°C for 5 hours. Cytop S-grade (100%; undiluted) annealed at 110°C for 5 hours. | Annealed at 80°C for 14 hours **after** gate evaporation. |
| 4 | N2200 Pristine 7 g/L (100% DCB) annealed at 110°C overnight (14 hours). | Cytop M-grade (3:1) annealed at 110°C for 5 hours. Cytop S-grade (1.5:1) annealed at 110°C for 5 hours. | Annealed at 80°C for 14 hours **after** gate evaporation. |

## III. RESULTS

### A. ELECTRICAL CHARACTERISTICS

**FIG. 2** shows the electrical device characteristics, and the mobility dependence with respect to the gate voltage for the fabrication conditions mentioned before. The on and off currents are of similar magnitude and do not depend on the annealing time, or the type of dielectric used. In fact, the worse performing device was the one where the gate was evaporated after the 80°C annealing step (red curve; recipe 2 in the Experimental Methods Section). This would indicate that the annealing step heals traps that are created during the gate evaporation. The corresponding output curves are shown in **FIG. S4**.

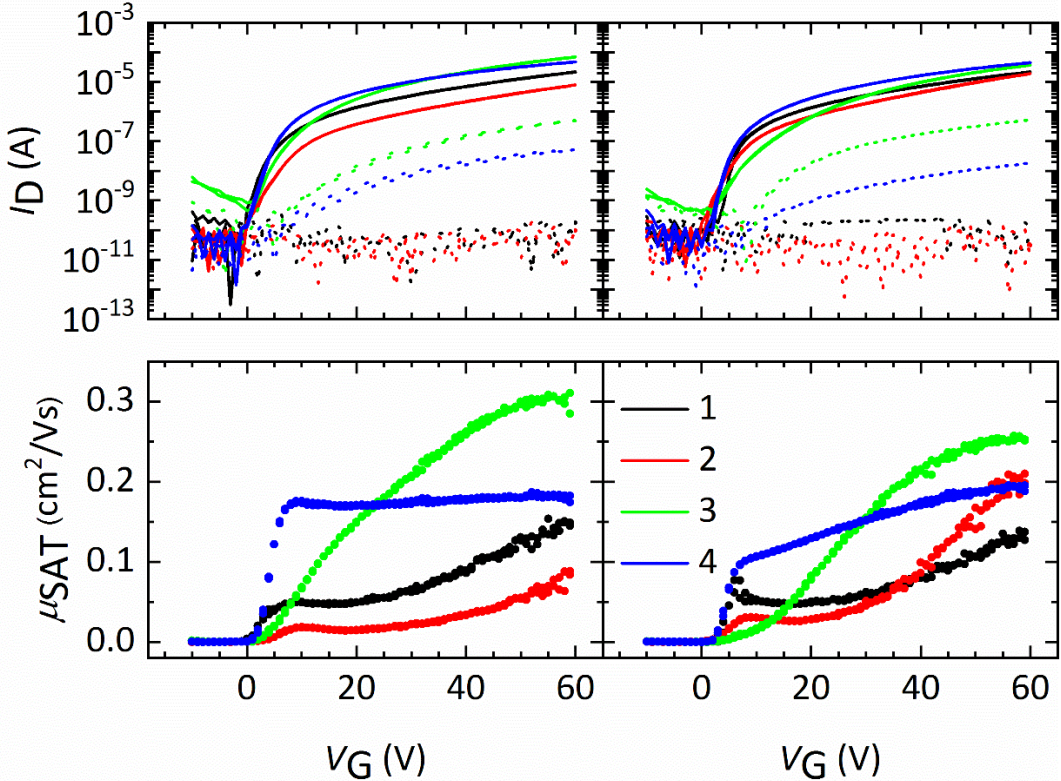

**FIG. 2.** Room temperature transfer characteristics and corresponding mobility curves measured in saturation ($V_D = 60V$) for N2200 devices fabricated with the Cytop-M (left panels) or the Cytop-S (right panels) dielectric. The continuous lines denote the drain current, and the dotted lines denote the leakage current. The black/red/green/blue curves correspond to the devices fabricated with recipe No. 1, 2, 3, and 4, respectively. The corresponding output curves are shown in **FIG. S4**.

Looking at the mobility dependence with respect to the gate voltage, there is a significant difference between the devices. The devices that were annealed for a short time (recipes 1 and 2) exhibit a power law dependence of the mobility with the gate voltage, as mentioned in the literature.[16] The origin of this power law dependence is believed to be a direct consequence of the shape of the density of states that is typical of organic semiconductors.[19,20,21] The literature suggests that the annealing temperature can be used to tune the size of the crystalline regions that are embedded in the amorphous matrix. In some previous studies it was found that higher annealing temperatures lead to larger crystalline regions, and to a stronger gate voltage dependence of the mobility. As-cast or mildly annealed samples were found to exhibit a mobility that is independent of the gate voltage.[16,22]

Our results convincingly show that these conclusions are not absolute, and that they depend sensitively on the details of the fabrication conditions. The devices in which the polymer was annealed for a short time above the boiling point, do exhibit the power law dependence that is reported in the literature. However, we have also found an improved, long-term annealing recipe (110°C for 14 hours) that improved performance further, and also returns to the less strongly gate voltage dependent characteristics that were previously observed in as-cast films (see **FIG. S3**). In this work, special care was taken to avoid all possible contaminations, e.g. from lab equipment such as pipettes or possible cross-contaminations in the glovebox.

Turning now to a discussion of the effect of end groups we were initially concerned about Cytop-M's silane end-groups being reactive with water, absorbing water, or forming other byproducts that compromise device performance.[14] In comparison, Cytop-S has a less reactive end group. For Cytop-S we optimized the fabrication to get films of similar thickness with Cytop-M. Somewhat surprisingly, we found the initial device performance and characteristics for both Cytop grades to be very similar. For both grades the devices annealed at 200°C for 15 minutes exhibit a strongly gate voltage dependent mobility with a positive curvature for gate voltages above $V_G > 20V$, while the devices made with long annealing times, including the pure Cytop-S films, exhibit a weaker gate voltage dependence with either a negative curvature above $V_G > 20V$ (recipe 3) or even a nearly gate voltage independent mobility (recipe 4). These findings suggest that the end group of the Cytop dielectric is not the dominant factor in determining the device performance of N2200.

## B. BIAS STRESS STABILITY

We then performed positive bias stress (PBS) stability tests on all the devices, to determine the effect of fabrication conditions on the operational stability, with the results displayed in **FIG. 3**. The gate and drain electrodes were biased at 60V for 6 hours, in dark conditions, and were left to recover for at least 3 hours. The devices were tested in a nitrogen atmosphere in a Belle Ltd. glovebox (<5 ppm O2). A transfer curve was measured every hour, and a corresponding spike appeared in the drain current, due to the partial recovery of the device during the measurement. The drain current was normalized with respect to its initial value at t=0.

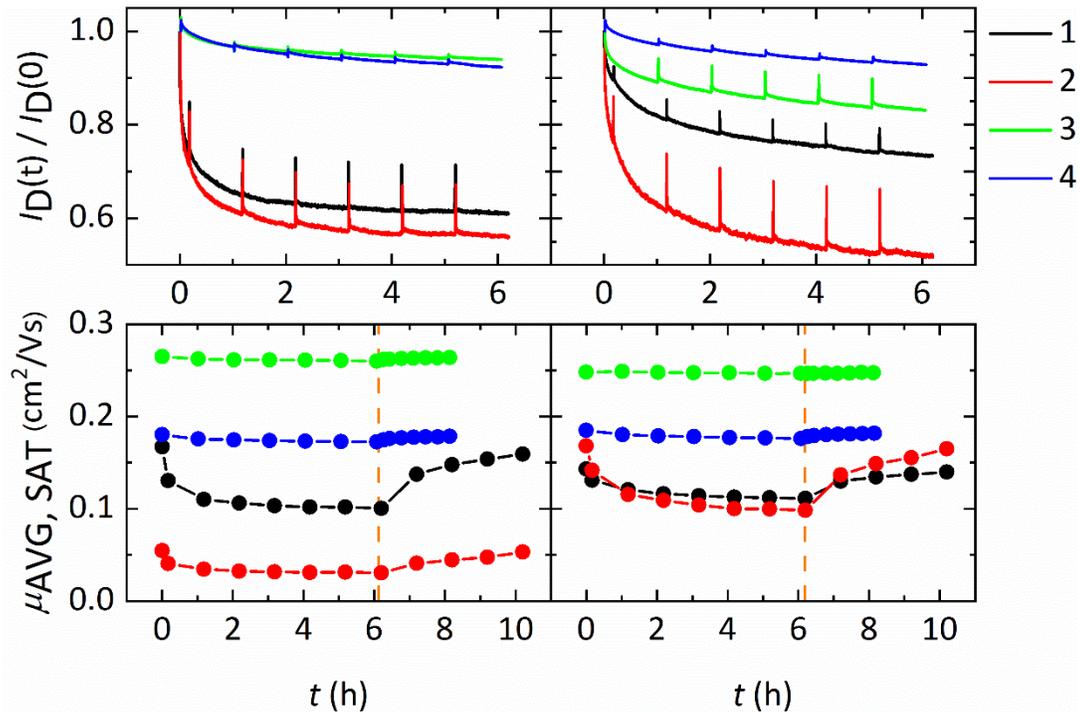

**FIG. 3.** A comparison of normalized drain current and saturation mobilities from positive bias stress (PBS) measurements for Cytop-M (left panels) or the Cytop-S (right panels). Long annealing times for both the polymer and the Cytop lead to high PBS stability. The vertical line separates the stress and recovery regimes. The average mobility was obtained by calculating the slope of the square root of the drain current in the voltage range between 40V and 60V, instead of calculating the point derivative. The black/red/green/blue curves correspond to the devices fabricated with recipe No. 1, 2, 3, and 4, respectively.

The top row of **FIG. 3** shows that the devices fabricated with short annealing times (recipes 1 and 2) exhibit rapid initial falloff of the drain current, whereas the devices annealed for long time scales only exhibit slight long-term degradation. This is somewhat different in Cytop-S, where the device that was annealed after the gate was evaporated (black curve, recipe 1) is significantly more stable than its counterpart made with Cytop-M.

The bottom row of **FIG. 3** shows that the devices fabricated with short annealing times (recipes 1 and 2) suffer from a reduction of the charge carrier mobility when stress is applied. When the device is left to recover, the mobility returns to its original value. This degradation effects are less pronounced for the devices made with Cytop-S. Conversely, the devices made with long annealing times exhibit a stable mobility, both in the stress and the recovery phase. This stability is independent of the type of dielectric used. The corresponding comparison of the threshold voltage shifts for these devices can be found in the SI and shows that $V_T$ shifts are not the dominant factor causing bias stress degradation.

The PBS results show that long annealing times for both the polymer and the Cytop lead to higher bias stress stability. The final to initial drain current ratios of the devices that were annealed for 15 minutes are respectively 61% (device annealed after the gate fabrication) and 56% (device annealed before the gate fabrication). Adopting the long annealing time fabrication increases these ratios to 94% (device with pure Cytop-M) and 92% (device with diluted Cytop-M).

For Cytop-S the values are similar to those of Cytop-M. The ratios of the devices that were annealed for 15 minutes are 73% (device annealed after the gate fabrication) and 52% (device annealed before the gate fabrication). Adopting the long annealing time fabrication increases these ratios to 83% (device with pure Cytop-S) and 93% (device with diluted Cytop-S). We can thus conclude that long annealing times are crucial for obtaining a PBS stable N2200 device, with the Cytop's end group being relatively unimportant in comparison.

## C. STRUCTURAL CHARACTERIZATION

We then examined whether the improved bias stress performance could be attributed to improved crystallinity, as was reported in previous studies.[11] We performed grazing incidence wide angle X-ray scattering measurements (GIWAXS) on N2200 films annealed at 200°C for 15 minutes (**FIG. 4**, left panel), and at 110°C for 15 hours (**FIG. 4**, right panel). The assignments of the diffraction peaks to indices were done in reference to the literature.[23,24]

The 2D pattern of the sample annealed at 200°C for a shorter time shows stronger diffraction peaks. It also has longer coherence lengths along the along the π–π stacking direction (51 Å) than the device annealed at 110°C for long annealing times (48 Å). Similarly, along the in-plane, lamellar stacking direction, the device annealed at 200°C for a shorter time exhibits longer coherence lengths (265 Å) than the device annealed at 110°C for long annealing times (222 Å).

Therefore, the sample that was annealed above the solvent's boiling point has slightly larger crystalline domains than the sample that was annealed below the boiling point for a longer time. We conclude that the improved bias stress stability in our N2200 TFTs obtained with the long-term annealing scheme cannot be traced to a higher crystallinity.

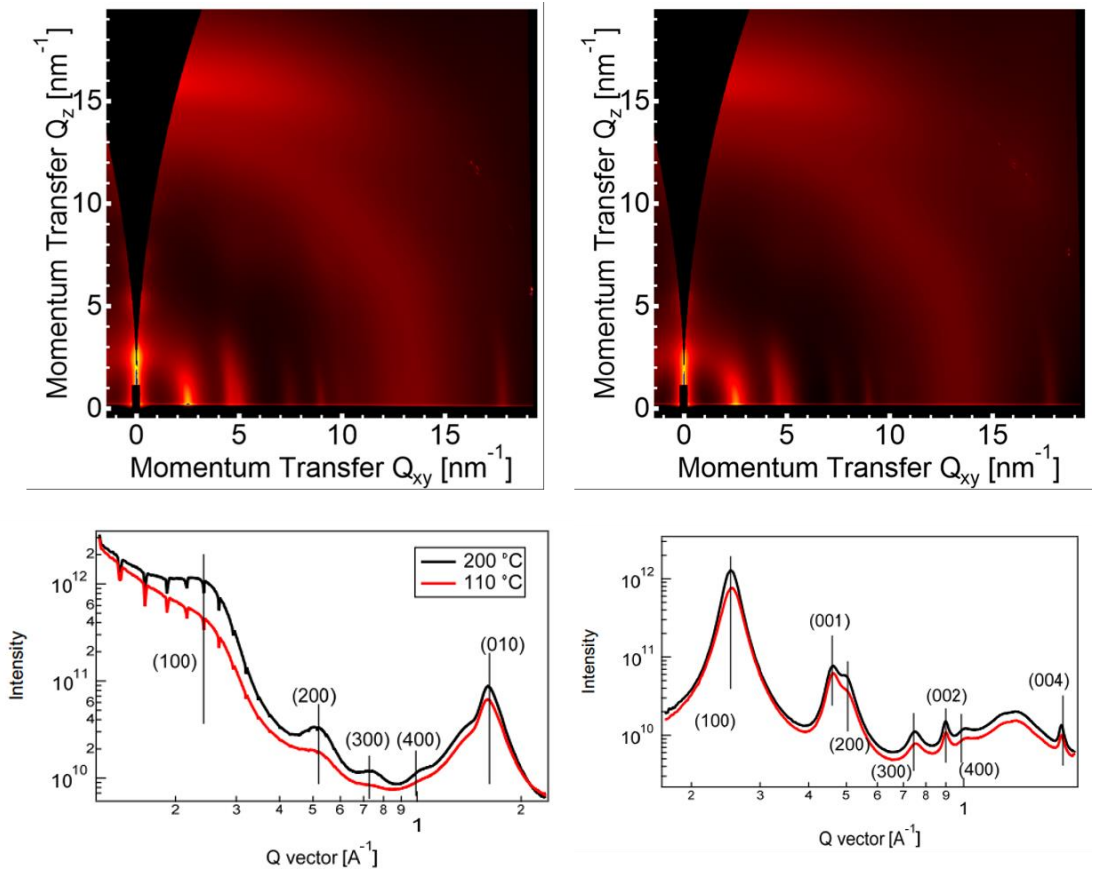

**FIG. 4.** GIWAXS results obtained on N2200 films annealed at 200°C for 15 minutes (left panel), and at 110°C for 15 hours (right panel). In plane (lamellar stacking; left panel) and out of plane (π-stacking; right panel) 1D GIWAXS profiles are shown. Coherence length and lattice parameters are extracted from these datasets and summarized in the table.

## IV. DISCUSSION AND CONCLUSIONS

In this work we have determined how the end groups of the gate dielectric and the choice of processing parameters impact performance and bias stress stability of the high mobility, n-type polymer N2200. We introduce Cytop-S as a well-performing gate dielectric and an alternative to Cytop-M with similar device performance and bias stress stability, i.e., the end groups of the Cytop do not influence initial device performance nor bias stress stability strongly. This is an important result as it broadens the range of end group functionalities that can be used in device applications. We caution that this result may be specific to N2200 and its molecular structure, which has long side chains that keep the accumulation layer away from the surface of the gate dielectric.[17] It is known that dipolar disorder is relatively short range.[25] This may not necessarily be the case for all polymers. It remains to be seen whether Cytop-S could improve charge transport performance in polymers with shorter side chains, where dipole-induced disorder due to the silane end group of Cytop-M could lead to more pronounced charge trapping.

Our study also provides important insight into some of the key factors that do influence bias stress stability under the inert atmosphere conditions investigated here. We have shown that annealing above the boiling point for short annealing times improves the crystallinity, but the higher degree of crystallinity is not associated with better bias stress stability. However, longer annealing times at low temperature improve bias stress stability significantly, even when annealing below the solvent's boiling point. When interpreting this result, it is important to remember that the degradation upon bias stressing under the inert atmosphere conditions used here is mainly a reduction in carrier mobility as opposed to threshold voltage. This means that it is not due to an increased fraction of deeply trapped charge carriers that do no longer contribute to charge transport – this would manifest itself mainly as a shift in the threshold voltage - but rather a change in the electronic structure at the interface that adversely affects all carriers and reduces their carrier mobility.

It is also important to remember that there appears to be a correlation between the gate voltage dependent mobility that is measured after the devices were fabricated and before the bias stress was applied and the degree of bias stress degradation. Devices with a more pronounced gate voltage dependent mobility degrade more strongly than devices with a gate voltage independent or more weakly gate voltage dependent mobility. This suggests that the cause of the energetic disorder in the electronic structure, that is responsible for the initial gate voltage dependence of the mobility, might also be responsible for the bias stress degradation. Whatever causes this energetic disorder is strongly reduced by long annealing of the polymer semiconductor and by long annealing of the full device. We are unable to identify the precise mechanism at this stage, but it appears possible that it could be associated with charged or polar impurities in the polymer films, such as water or residual polar solvent molecules. The application of a bias might bring these impurities closer to the interface or give more time for charge carriers in the N2200 to polarize these impurities, in both cases the expected effect would be an increased energetic disorder and reduced mobility. It is also possible that the disorder may be associated with states created during the gate evaporation, for example, due to diffusion of metal atoms into the gate dielectric. Long-term annealing might be able to passivate such states, provided that the annealing step is performed after the gate evaporation. If the annealing precedes the gate evaporation then the device performs poorly.

Identification of the species and states involved in the degradation is difficult due to lack of analytic techniques with sufficient sensitivity to the small concentrations involved and this goes beyond the scope of our present study. However, what we can state firmly from our study is that the gate voltage dependent mobility and faster bias stress degradation observed in devices annealed at high temperature for short times are not related to a reduced degree of crystallinity of the polymer semiconductor suggesting that the degree of crystallinity is not a key factor involved in bias stress degradation and that the absence of end groups of the Cytop dielectric do not appear to have a significant effect on performance and bias stress stability of N2200.

## V. EXPERIMENTAL METHODS

**Recipe 1:** A 7 g/L solution of N2200 and DCB, kept in a nitrogen glovebox, and was pre-heated at 100°C overnight prior to device fabrication leading to a fully dissolved polymer solution. Corning 1737F glass substrates with Cr/Au (4/20 nm) interdigitated source-drain electrodes (W = 1000 um, L = 20 um) were pre-heated at 100°C for 2 minutes directly before fabrication. With the glovebox purging on, the organic semiconductor was spin coated at 2000 rpm for 60 seconds (ramp: 3 seconds) and annealed on a hotplate at 100°C for 10 seconds, to remove most of the solvent. After all the samples were fabricated, the temperature was re-adjusted to 200°C, with a hotplate ramp of 450°C/h. The samples were left at 200°C for 15 minutes.[16] The hotplate heater was then switched off and the samples taken off when the temperature reached at 80°C. Cytop M-grade (CTL-809M) and Cytop S-grade (CTX809-SP2) were purchased from AGC Inc. Dielectric solutions of Cytop-M (Cytop to solvent ratio 3:1 v/v) and Cytop-S (Cytop to solvent ratio 1.5:1 v/v) were mixed the day before and were shaken vigorously to mix the solutions prior to deposition. In the case of Cytop-M, spin-coating was carried out in two steps: first at 500 rpm, for 3 seconds (ramp: 1 sec), and then at 2000 rpm for 30 seconds (ramp: 3 sec). It was annealed at 90°C for 20 minutes. This resulted in a thickness of 550 ± 20 nm, as measured with a Dektak XT profilometer. In the case of Cytop-S, spin-coating was carried out in a single step, at 1200 rpm for 60 seconds (ramp: 3 sec). It was annealed at 90°C for 20 minutes. This resulted in a thickness of 490 ± 20 nm, as measured with a Dektak XT profilometer. Cytop-M was drop cast on the sample surface with a pipette in a zig-zag fashion, making sure to cover the entire surface of the sample before the spin-coating commenced. Cytop-S was deposited on the center of sample, allowing it to spread outwards and cover the entire sample.

An MBraun MB-20-G glovebox (kept at <1 ppm O2, <1 ppm H2O) and an MBraun HPL 150 ECO metal hotplate were used throughout the entire pre- and fabrication process. The annealing was carried out with the purging activated and the laminar flow deactivated. Samples were always being transferred in sealed nitrogen filled pass tubes, or otherwise stored in a nitrogen atmosphere in a Belle Ltd. glovebox (<5 ppm O2) when not being processed. An evaporation of 40 nm thick Al gates was carried out in an EVOVAC Angstrom Engineering evaporator. The evaporator shields were covered with Al foil before each evaporation to prevent cross-contamination. The evaporation was performed immediately after a chamber pressure of $1.65*10^{-6}$ Torr was achieved. Deposition rates of 0.19-0.22 A/s were used for the first 10 nm and 0.89-1.18 A/s for the remaining 30 nm with the process pressure varying between $1.65*10^{-6}$ at the beginning to $9.84*10^{-7}$ Torr at the end of the evaporation. The resulting gates were uniform of good quality when inspected with an optical microscope. The devices were subsequently annealed overnight (14 hours) at 80°C on a metal hotplate inside the fabrication glovebox.

**Recipe 2:** The process follows the one described for the 1st recipe; however, with the overnight (14 hours) 80°C device annealing step performed prior to Al gate evaporation. Deposition rates of 0.19-0.22 A/s were used for the first 10 nm and ~1 A/s for the remaining 30 nm.

**Recipe 3:** A modified version of the recipe in ref. 17 was used. The fundamental differences to the above recipes involve the use of Cytop with no solvent, 9000 rpm spin-coating for 90 seconds, annealing without purging, a 14h long annealing step at 110°C directly following the N2200 spin coating, and a 5h long annealing step at 110°C following the Cytop dielectric spin coating. The thicknesses for pure Cytop-M and Cytop-S were $310 \pm 25$ nm, and $570 \pm 50$ nm respectively, as measured with a Dektak XT profilometer. It is worth noting that further annealing at 80°C on the whole device, i.e. after the whole fabrication is finished, does not improve the performance of these devices much further. Deposition rates of ~0.5 A/s were used for the first 7 nm and ~2 A/s for the remaining 13 nm.

**Recipe 4:** The same recipe as that for recipe 3 with a standard dilution of Cytop to solvent (a ratio of 3:1) and annealing carried out on a ceramic hotplate, the latter of which could translate into a variation in the annealing temperature compared to the metal hotplates used for the other batches. The ceramic hotplate had approximately double the standard deviation of the setpoint temperature of the metallic hotplate when measured in their most central points, i.e. within the heating elements (~6% vs. ~3%) Deposition rates of ~0.4 A/s were used for the first 10 nm and ~1 A/s for the remaining 10 nm.

The N2200 OFETs were characterized in a N2 glovebox using an Agilent 4155B Semiconductor Parameter Analyzer. Bias stress measurements were performed in the dark.

GIWAXS was performed at the SAXS/WAXS beamline at the Australian Synchrotron.[26] An X-ray energy of 12.0 keV was used, with scattering recorded on a Dectris PILATUS3 X 2M in-vacuum detector placed ~ 69 cm downstream from the sample. The entire flight path was in vacuum. Scattering patterns were carefully measured as a function of incident angle (down to a resolution of 0.01°) with the reported images taken close to the critical angle where scattering intensity was maximized. Further details of data analysis and processing can be found elsewhere.[27]

**SUPPLEMENTARY MATERIAL**

Additional fabrication and electrical data are presented in a table and figures in the supplementary material.

**ACKNOWLEDGEMENTS**

D.S. acknowledges support by the Sensor CDT and the Engineering and Physical Sciences Research Council (Grant No. EP/L015889/1). L.J.S. and D.V. acknowledge support by the ERC Synergy Grant SC2 (Grant No. 610115). U.K. acknowledges funding from the EPSRC (Grant No. EP/R031894/1). M.N. acknowledges financial support from the European Commission through a Marie-Curie Individual Fellowship (EC Grant Agreement Number: 747461). This work was performed in part at the SAXS/WAXS beamline at the Australian Synchrotron, part of ANSTO. The authors acknowledge support by the Henry Royce Institute: Cambridge Equipment: EP/P024947/1. The authors thank AGC Inc. for supplying Cytop-S for this study. D.V. thanks Paul Barton of AGC Inc. for his continuous support during this project. The authors declare no conflict of interest.

**DATA AVAILABILITY**

The data that support the findings of this study are available from the corresponding author upon reasonable request.

# The effect of the dielectric end groups on the positive bias stress stability of N2200 organic field effect transistors – supplementary information

D. Simatos[1,b], L. J. Spalek[1,a),b], U. Kraft[1], M. Nikolka[1], X. Jiao[2], C. R. McNeill[2], D. Venkateshvaran[1,a], H. Sirringhaus[1,a]

[1] Cavendish Laboratory, University of Cambridge, J.J. Thomson Avenue, Cambridge CB3 0HE, UK

[2] Department of Materials Science and Engineering, Monash University, Wellington Road, Clayton, VIC 3800, Australia

a) Authors to whom correspondence should be addressed: ljs51@cam.ac.uk, dv246@cam.ac.uk, hs220@cam.ac.uk

b) D. Simatos and L. J. Spalek contributed equally to this work.

The effect of post-fabrication annealing with residual solvent (the photo below obtained for devices fabricated using recipe 1). Post-fabrication annealing N2200 OFETs at 80°C overnight (recipe 1) on a metal hot plate leads to cracking of the Al gates, possibly due to the Cytop solvent escaping the film. The effect does not appear when annealing devices post fabrication which were subjected to the additional step of very long Cytop annealing times prior to gate evaporation (recipe 2). It is worth noting that the damaged gates did not degrade the device performance.

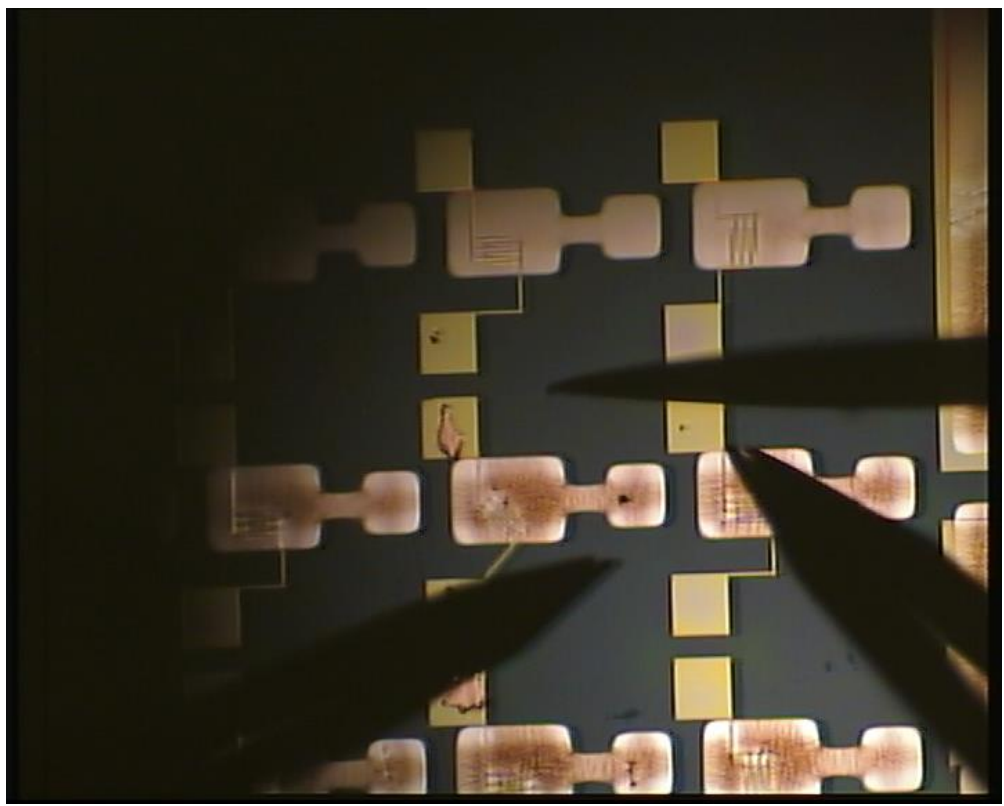

**FIG. S1.** The effect of post-fabrication annealing with residual solvent. Post-fabrication annealing N2200 OFETs at 80°C overnight on a metal hot plate leads to the cracking of the Al gates, due to the Cytop solvent escaping the film. The effect does not appear when annealing post-fabrication annealed devices with long Cytop annealing times.

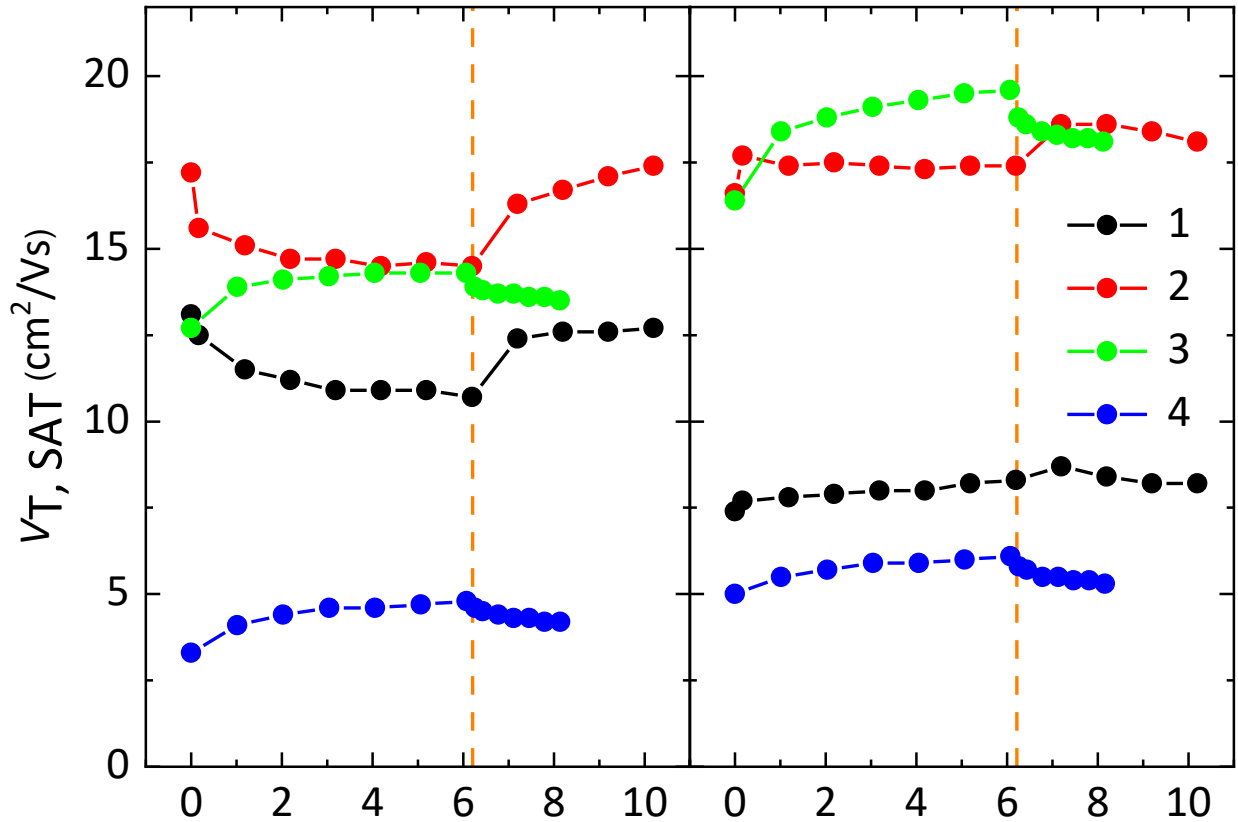

**FIG. S2.** A comparison of saturation threshold voltages from positive bias stress (PBS) measurements. Long annealing times for both the polymer and the Cytop lead to smaller voltage shifts. The devices made with Cytop-S were much more stable than those made with Cytop-M. The vertical line separates the stress and recovery regimes. The linear fit was performed by calculating the slope of the square root of the drain current in the voltage range between 40V and 60V. The black, red, green, blue curves correspond to the devices fabricated with recipe No. 1, 2. 3, and 4, respectively.

**FIG. S2** shows that the devices fabricated with short annealing times (recipes 1 and 2) exhibit shifts of the threshold voltage towards the negative voltages. The threshold voltages recover to their original values when the stress is lifted. The devices fabricated with long annealing times (recipes 3 and 4) are more stable. They exhibit slightly increasing threshold voltages during bias stress, and recover when the stress is lifted. The short-annealed devices fabricated with Cytop-S are more stable than their Cytop-M counterparts. Their threshold voltage also shifts to a different direction: while the devices made with Cytop-M have threshold voltages that drop to more negative values, the devices with Cytop-S exhibit slightly increasing threshold voltages during bias stress. We therefore conclude that the directions of the threshold voltage shifts are partly attributed to the end groups of the dielectric material used.

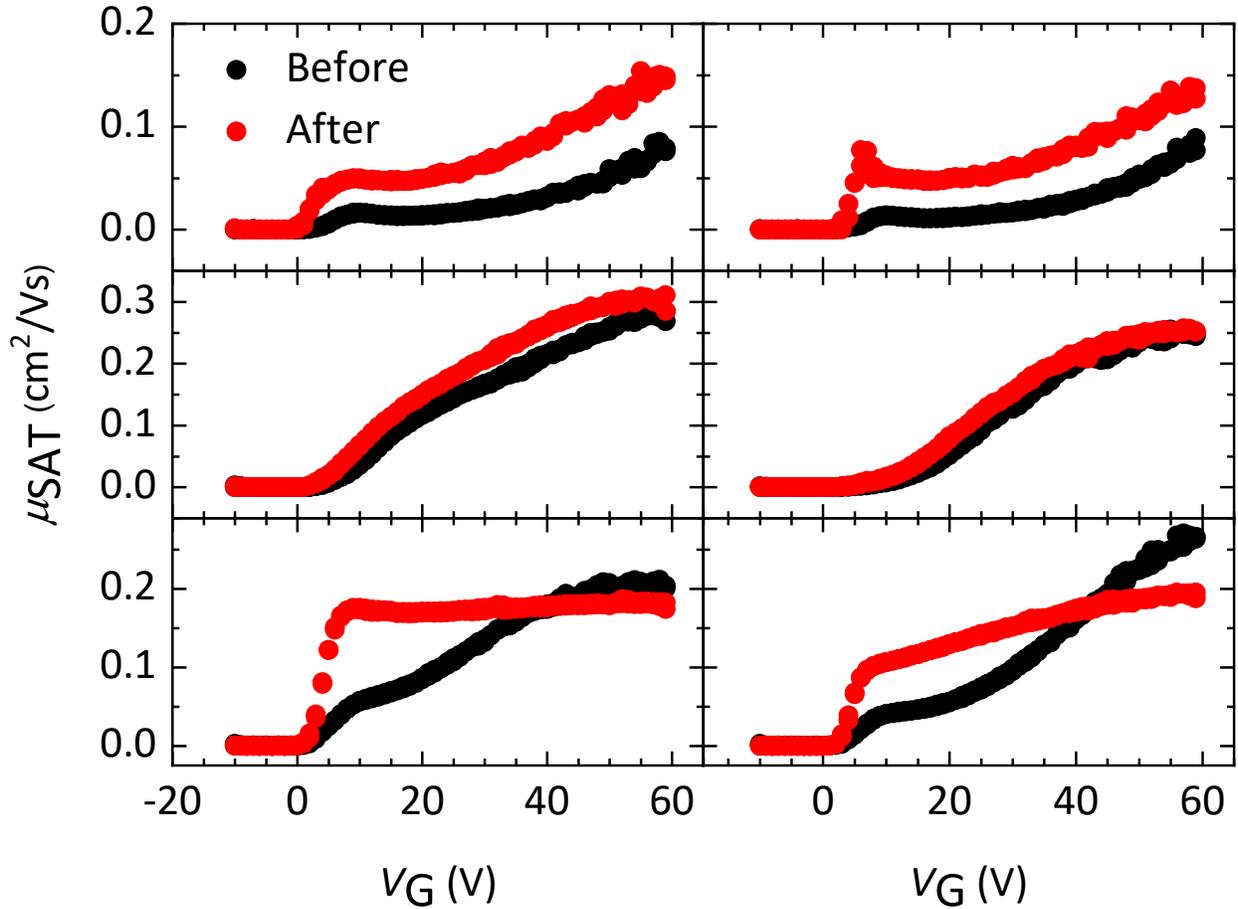

**FIG. S3.** shows the mobility dependence with respect to the gate voltage for three of the four recipes that we examined, before (black curve) and after (red curve) the post-fabrication annealing step. Recipes 1, 3, and 4 are depicted in rows 1, 2, and 3 respectively. The first column shows the devices made with Cytop-M, and the second column those made with Cytop-S. The device fabricated with short annealing times and annealed before the gate was evaporated (recipe 2) is not included because the device could not be measured before the annealing step, since the gate had not yet been fabricated.

The device fabricated with short annealing times and annealed after the gate was evaporated (recipe 1) confirms the power law dependence that is reported in the literature.[16] However, the devices fabricated with long annealing times (recipes 3 and 4) do not follow this trend. The device made with pure CYTOP-M exhibits a linearly dependent mobility, whereas the device made with the standard diluted CYTOP-M recipe has a mobility that is independent of the gate voltage, a trait that the literature assigns to as-cast films.[16,22] We can then conclude that the gate voltage dependence of the mobility is not necessarily correlated to as-cast films.

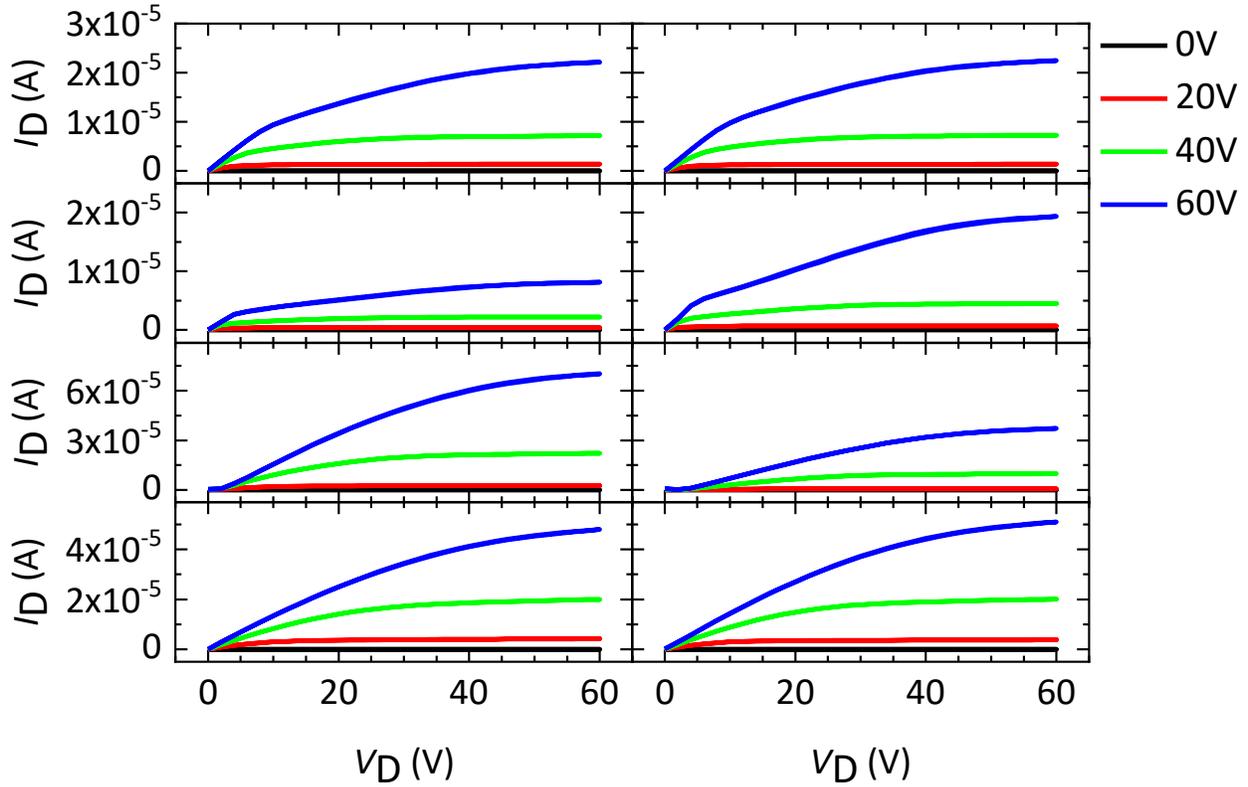

**FIG. S4.** Room temperature output curves for N2200 devices fabricated with the Cytop-M (left panels) or the Cytop-S (right panels) dielectric. The four rows correspond to recipes 1, 2, 3, and 4, respectively. The devices fabricated with short annealing times (recipes 1 and 2) exhibit kinks, which are not present in the devices fabricated with long annealing times (recipes 3 and 4).

**Table SI.** Table comparing the outcomes of fabrication recipes used in this work with the literature.

|  | N2200 anneal | Crystallinity | Cytop anneal | Post-anneal | μ vs Vg | PBS |
|---|---|---|---|---|---|---|
| This work, recipe 1 | Short (15 minutes), above boiling point | more | Short (20 minutes) | After gate | Power law | degraded |
| This work, recipe 2 | Short (15 minutes), above boiling point | more | Short (20 minutes) | *Before* gate | Power law | degraded |
| This work, recipe 3 | Long (14 hours), below boiling point | *less* | Medium (5 hours), ***pure Cytop*** | After gate | Linear | improved |
| This work, recipe 4 | Long (14 hours), below boiling point | less | Medium (5 hours) | After gate | independent of Vg | improved |
| [16] Statz et al | Short (15 minutes), above boiling point | more | PMMA | / | Power law | |
| [16] Statz et al | "as cast" (vacuum dried) | less | PMMA | / | Approximately independent | |
| [17] Caironi | Long (14 hours), below boiling point | / | Medium (5 hours) + PMMA + PS | / | | |